# Microstructural engineering of medium entropy NiCo(CrAl) alloy for enhanced room and high-temperature mechanical properties

Nithin Baler[*], Abdulla Samin M V, Akshat Godha, Surendra Kumar Makineni[*]

Department of Materials Engineering

Indian Institute of Science, Bangalore 560012, India

[*]Corresponding Authors: nithinbaler@iisc.ac.in, skmakineni@iisc.ac.in

**Abstract**

This work demonstrates the development of a strong and ductile medium entropy alloy by employing conventional alloying and thermomechanical processing to induce partial recrystallization (PR) and precipitation strengthening in the microstructure. The combined usage of electron microscopy and atom probe tomography reveals the sequence of microstructural evolution during the process. First, the cold working of homogenized alloy resulted in a highly deformed microstructure. On annealing at 700°C, B2 ordered precipitates heterogeneously nucleate on the highly misoriented sites. These B2 promotes particle stimulated nucleation (PSN) of new recrystallized strain-free grains. The migration of recrystallized grain boundaries leads to discontinuous precipitation of $L1_2$ ordered regions in highly dense lamellae structures. Atomic-scale compositional analysis reveals a significant amount of Ni confined to the GB regions between B2 and $L1_2$ precipitates, indicating Ni as a rate-controlling element for coarsening the microstructure. On 20 hours of annealing, the alloy comprises a composite microstructure of soft recrystallized and hard non-recrystallized zones, B2 particles at the grain boundaries (GBs), and coherent $L1_2$ precipitates inside the grains. The B2 pins the GB movement during recrystallization while the latter provides high strength. The microstructure results in a 0.2% yield stress (YS) value of 1030 MPa with 32% elongation at ambient temperature and retains up to 910 MPa at 670°C. Also, it shows exceptional microstructural stability at 700 °C and resistance to deformation at high temperatures up to 770°C. Examination of deformed microstructure reveals excessive twinning, formation of stacking faults, shearing of $L1_2$ precipitates, and accumulation of dislocations at around the B2 precipitates and GBs attributed to high strain hardening of the alloy.

# 1. Introduction

Design of single-phase alloys by increasing the configurational entropy, i.e., by mixing the multiple principal elements in equimolar or near equimolar ratios, showed a great promise with excellent mechanical properties at room and cryogenic temperatures [1–3]. The properties are directly related to the high configurational entropy and sluggish atomic diffusion kinetics with large lattice distortion [4]. The former stabilizes the single-phase relative to the multiphase structure while the latter resists dislocation motion on loading. In particular, the model face-centered-cubic (fcc) high entropy alloy (HEA) "FeCoNiCrMn" shows a substantial improvement of yield strength, ductility, and toughness as the temperature reduces up to cryogenic conditions [5]. However, at room temperature, the alloy yields at a very low stress (0.2% YS) value of 200 MPa, and by reducing the grain size (from 155 μm to 4.4 μm), the 0.2% YS increases up to only 350 MPa [6,7]. As the temperature increases, the 0.2% YS drops drastically, for example, by ~30% at 400°C [8]. Later, strengthening by dispersion of coherent $L1_2$ precipitates by adding Ti and Al to a similar single-phase fcc FeCoNiCr alloy was demonstrated that increased the 0.2% YS up to 645 MPa at room temperature with ~ 39% plasticity [9]. The microstructure also contains an additional $L2_1$ Heusler-like phase, mainly at the GBs. Here, the strengthening $L1_2$ precipitates are only stable up to 550°C, while at higher temperatures, the precipitates decompose, and ordered B2 precipitates appear inside the grains and at the GBs [10]. Cu addition improved the stability of $L1_2$ precipitates up to 700°C (for 50 hours), where Cu-rich clusters in the matrix act as heterogeneous nucleation sites for the $L1_2$ precipitates [11]. But at higher temperatures, Cu-rich needle shape precipitates dominate the microstructure. Hence, these alloys might have limitations to be used at such high temperatures. The main reason for the $L1_2$ precipitate instability is the presence of Fe in the alloys, which will be discussed in detail later.

On the other hand, NiCoCr alloy (medium entropy configuration) shows a better combination of mechanical properties (yield strength, ductility, and fracture toughness) than quaternary and quinary alloys at room and cryogenic temperatures. This was attributed to the reduction of stacking fault energy (by ~ 25%), which is proposed to resist cross-slip and promote nano-twinning as the strain accumulator during deformation [12]. As for MEA, the feasibility of forming $L1_2$ precipitates has been recently exploited by adding Al and Ti [13–15]. An increase in Y.S. by ~ 70% (up to 750 MPa) is reported [15]. However, studies on their high-temperature properties are limited [14,16]. The H/MEAs were also shown to be highly sensitive towards partial recrystallization (PR) that led to the formation of hierarchical microstructures and offered a new thermo-mechanical processing platform for designing alloys overcoming the strength-ductility tradeoff at room temperature [17,18].

In this work, we demonstrate the application of PR to an MEA NiCo(Cr$_{0.72}$Al$_{0.28}$) that resulted in tensile 0.2% YS of 1030 MPa with the ductility of 32% at room temperature and retained compressive 0.2% YS of 910 MPa at 670°. The obtained microstructure is stable at 700°C without any microstructural and mechanical degradation up to 500 hours. The microstructural features were identified and discussed in correlation with the obtained properties. We also propose new solute diffusional mechanisms and sheds light on the possible rate-limiting solutes responsible for the high-temperature microstructural stability of the present investigated alloy.

## 2. Experimental

*2.1 Alloy preparation*

Alloy ingots with nominal compositions (all in at.%) of 33.33Ni-33.33Co-33.33Cr (here onwards NCC) and 33.33Ni-33.33Co-24.33Cr-9Al (here onwards NCCA) were melted in the form of 50 g buttons using a vacuum arc melting unit under argon atmosphere. The constituent elements have 99.99% purity. The alloys were remelted 5 to 6 times by flipping after each cycle of melting for homogeneity. The buttons were subsequently cast in the form of a slab (70 mm x 20 mm x 10 mm) using a water-cooled copper split mold in a vacuum casting unit under an argon atmosphere.

*2.2 Thermomechanical processing*

The cast slabs were homogenized (H) at 1200 °C for 20 hours in a tubular furnace maintained under a vacuum of 10$^{-5}$ mbar followed by water quenching. The homogenized samples were cold-rolled (C) to a reduction ratio of 70 % and subsequently annealed (A) at a temperature of 700 °C for 20 hours, followed by water quenching. This thermomechanical heat treatment will be called HCA (Homogenization (H) + cold working (C) + annealing (A)) in the rest of the manuscript. The microstructural stability was also evaluated by annealing at 700°C for up to 500 hours.

*2.3 Microstructural characterization*

Identification of phases for the annealed samples was conducted using an X-ray diffractometer (XRD, Rigaku) operated at room temperature with Cu Kα source equipped with Johansson optics that eliminates the Kα2 and Kβ components. The phase transformation temperatures were determined in a differential scanning calorimeter (DSC, STA449F3 NETSCH). DSC samples weighing 50 mg were cut from the alloy after HCA thermomechanical processing and heated at a 10 K/min rate under argon atmosphere till 1450°C and cooled at the same rate to room temperature.

After each experimental step, the microstructures were characterized by using a scanning electron microscope (SEM, Helios Nanolab) equipped with a field-emission-gun source (FEG). Electron backscattered diffraction (EBSD) patterns were recorded in Gemini SEM450 (Carl Zeiss) equipped with a velocity detector using a step size of 0.1µm. For SEM and EBSD analysis, samples were prepared following a standard metallography technique using Si grit papers and final polishing in a Vibro-polisher (VibroMet Buehler made) with colloidal silica suspension of 200 nm particle size. The high temperature deformed samples were also investigated by controlled electron channeling contrast image (cECCI) using a Zeiss SEM (Carl Zeiss) equipped with a Gemini field emission gun electron column operated at 30 kV with a probe current of 5 nA. The working distance was kept at ~ 6 mm during imaging. First, the deformed region of interest (ROI) on the bulk sample was identified, and the crystallographic orientation of the region was measured by EBSD. A pole figure was generated for the ROI using orientation imaging microscopy software (OIM). The exact orientation was generated in a single crystal software to access the Kikuchi pattern for the particular orientation directly. Using single-crystal software, we obtained the stage tilt and rotation values required for the ROI to be in two-beam conditions. The microscope stage was aligned according to the measured values, and imaging was done using a backscattered electron detector (BSE).

Electron diffraction analysis was performed using a transmission electron microscope (TEM, Tecnai T20) operated at 200 kV. High-resolution imaging was done in an aberration-corrected TEM in STEM mode (Titan Themis) operated at 300 kV. Atomic-scale compositional analysis was performed by atom probe tomography (APT) in LEAP-5000XHR (Cameca instruments) equipped with a reflectron. An ultraviolet picosecond pulse laser with a pulse repetition rate of 125 kHz and pulse energy of 55 pJ was used for field evaporation. The base temperature of needle specimens was maintained at 60 K with a target detection rate of 5 ions per 1000 pulses. The APT data analysis and reconstruction were performed using the IVAS$^{TM}$ 3.8.4 software package. TEM and APT specimens were prepared by standard site-specific in-situ-lift-out protocols [19] using a dual-beam SEM/focused ion beam (FIB) instrument (ThermoFisher Scios) operated at 30 kV. APT specimens from the deformed regions were also prepared directly from the TEM lamella after the TEM experiments from the exact location. A final cleaning of TEM lamella and the APT needles was carried out at 2 kV with an 8 pA current to remove the damaged regions caused during prior Ga$^+$ ion milling.

*2.4 Mechanical properties*

Room temperature tensile and high temperature compressive mechanical properties were evaluated using an Instron 5967 UTM (Ultimate Tensile Machine) loaded at a constant strain rate of $10^{-3}$ s$^{-1}$. Dog-bone shape flat specimens were cut from the alloys after HCA thermomechanical heat treatment for tensile testing, per the ASTM standard E8. Cylindrical samples of 3 mm diameter and 4.5 mm height (heigh/diameter ratio = 1.5) were also cut from the alloy for a high-temperature compression test. The hardness values were measured using a Vickers microhardness tester (FutureTech, FM-800) by using a load of 0.5 kg.

### 3. Results

*3.1 Mechanical Properties*

Figure 1(a) shows uniaxial tensile test engineering stress-strain curves for both NiCoCr (NCC) and NiCoCrAl (NCCA) MEA alloys after HCA thermomechanical heat treatment. The tensile yield strength (0.2% YS) and ultimate tensile strength (UTS) for NCC alloy are measured to be 518 MPa and 860 MPa, respectively, with the elongation to failure ~ 48 %. In comparison, the NCCA alloy show 0.2% YS and UTS of 1030 MPa (higher by ~99%) and 1320 MPa (higher by ~53%), maintaining a ~32% elongation to failure. Figure 1(b) shows an Ashby plot between UTS and elongation that compares the values of NCCA alloy (red star) with Fe containing NiCoCrMn based HEA and other critical engineering alloys. It demonstrates a unique place with a promising combination of mechanical properties at room temperature. Further, the high-temperature properties of NCCA alloy are evaluated using compression tests up to 770°C. Figure 1(c) shows a comparison of 0.2% YS values vs. temperature for NCCA alloy with few earlier reported values for HEA and some commercial Ni-based and Co-based superalloys. The 0.2% YS for NCCA shows decreasing tendency as the temperature increases and maintains a high value of ~ 910 MPa at 670 °C, comparable to Ni-base MAR-M-247 superalloy and superior to even some of the refractory HEAs. At 770°C the value reduces to 600 MPa. For microstructural stability at high temperature, the NCCA alloy was annealed at 700°C up to 500 hours, and hardness was taken at room temperature after different time intervals. The selection of the temperature to examine stability was based on DSC experiments which will be discussed later. Figure 1(d) shows the hardness variation with the annealing time, indicating a negligible change in the values. Hence, the NCCA alloy show promising mechanical properties and microstructural stability both at room and high temperatures.

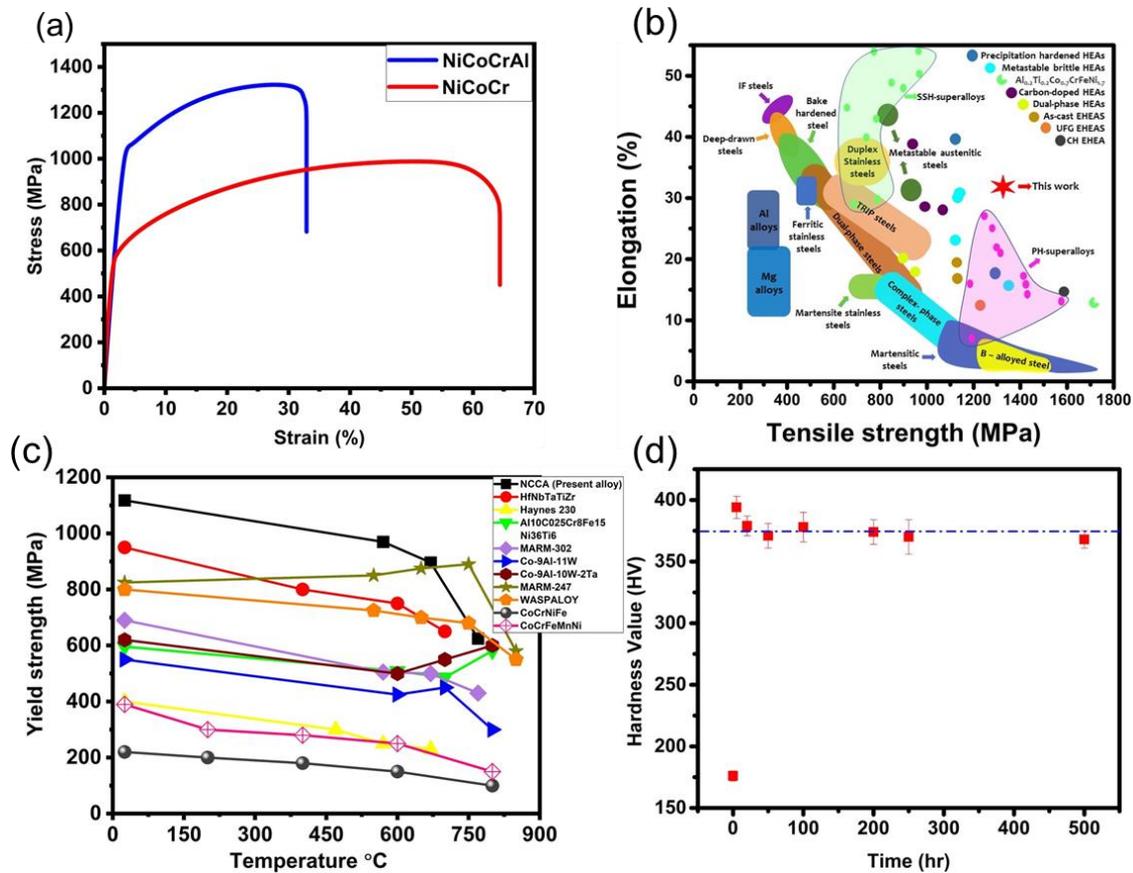

*Figure 1: (a) Comparison of engineering tensile test plots of NCC and NCCA alloys. (b) Ashby plot between tensile strength and elongation Comparing the position of NCCA (red star) alloy with HEAs, superalloys, and other engineering alloys [20–28]. (c) Comparison of 0.2% YS vs. temperature for NCCA alloy with some of the HEAs (bcc and fcc based) and superalloys [29–36]. (d) Variation of hardness value of NCCA alloy with annealing time at 700°C.*

## 3.2 Microstructural analysis

Now, we detail the microstructure features of NCCA alloy in correlation with the obtained properties. Fig. 2(a & b) compares secondary electron (SE) micrographs of NCC and NCCA alloys after HCA thermomechanical heat treatment. NCC alloy is fully recrystallized with fine equiaxed grains of average size ~6 µm. Inset in 2(a) shows the electron backscattered diffraction (EBSD) inverse pole figure (IPF) map from a region of the alloy. In contrast, NCCA alloy is partially recrystallized (PR); see 2(b) inset for the IPF map. It contains ~70% soft recrystallized zones of fine grains (1.5 to 2 µm) and hard non-recrystallized zones of coarse grains (several tens of micron sizes (marked in red arrows)) with an average size of ~ 3.4 µm.

Another inset in fig.2(b) shows a higher magnification micrograph that reveals two kinds of second phase precipitates formed along the grain boundaries (GBs) and interior of the recrystallized grains.

X-ray diffraction (figure 2(c)) analysis for the NCCA alloy after homogenization (H) confirms the presence of a single fcc phase throughout the microstructure. However, after HCA, additional diffraction peaks correspond to B2 ordered structure appear. From the microstructure in figure 2(b), the B2 peaks might be attributed to the precipitates at the GBs.

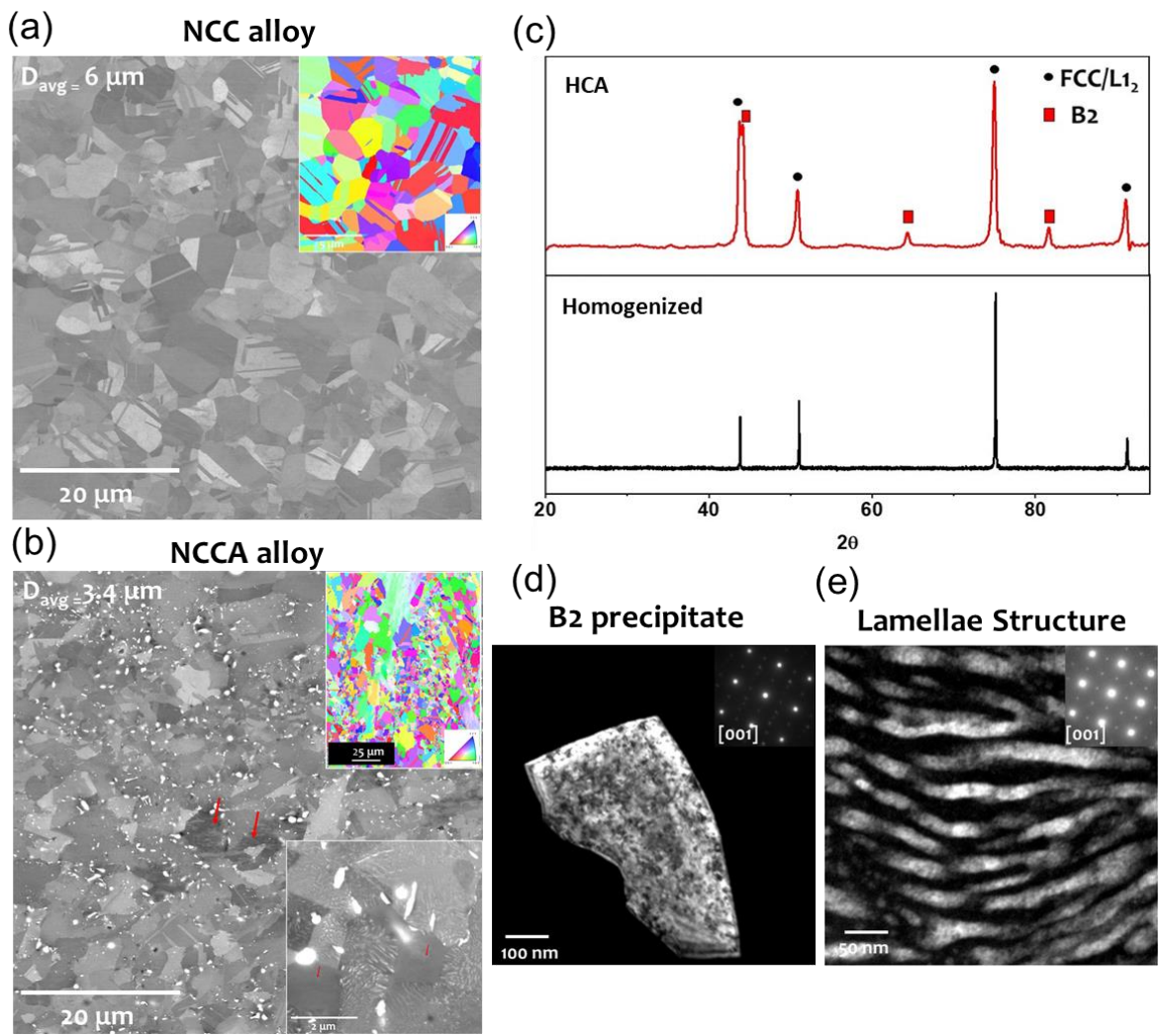

*Figure 2:* SEM images for (a) NCC and (b) NCCA alloys after HCA thermomechanical heat treatment and their corresponding EBSD IPF maps as insets. (c) Comparison of XRD patterns of NCCA alloy after homogenization (H) and after HCA thermomechanical heat treatment. Darkfield (DF) images highlighting (d) a B2 ordered precipitate located at a GB triple junction and (e) $L1_2$ ordered precipitates from a lamellae region inside of a recrystallized grain.

The fine precipitates inside the recrystallized grains are in the lamellae form that is typical of discontinuous precipitation. Note that the lamellae precipitates are not present in non-recrystallization zones (see inset figure in 2(b)), indicating their discontinuous formation during the migration of GBs on recrystallization [37]. The morphology of the lamellae structure shows the spacing is relatively higher adjacent to the GBs than the interior of the grains.

The structure and composition of the evolved precipitates are determined using TEM and APT. Fig. 2(d) shows a darkfield (DF) micrograph of a precipitate that is formed at a GB triple junction. The corresponding diffraction pattern is shown as an inset and can be indexed as the [001] pattern of the B2 ordered crystal structure. Similarly, the diffraction pattern from the interior of a recrystallized grain (see inset figure 2(e)) indicates the presence of additional superlattice reflections along with the primary fcc reflections in the [001] oriented pattern. These reflections are attributed to $L1_2$ ordering, and the darkfield from one of the superlattice reflections highlights the ordered regions in the form of lamellae structure, figure 2(e). The average spacing between the adjacent ordered regions is ~29 nm.

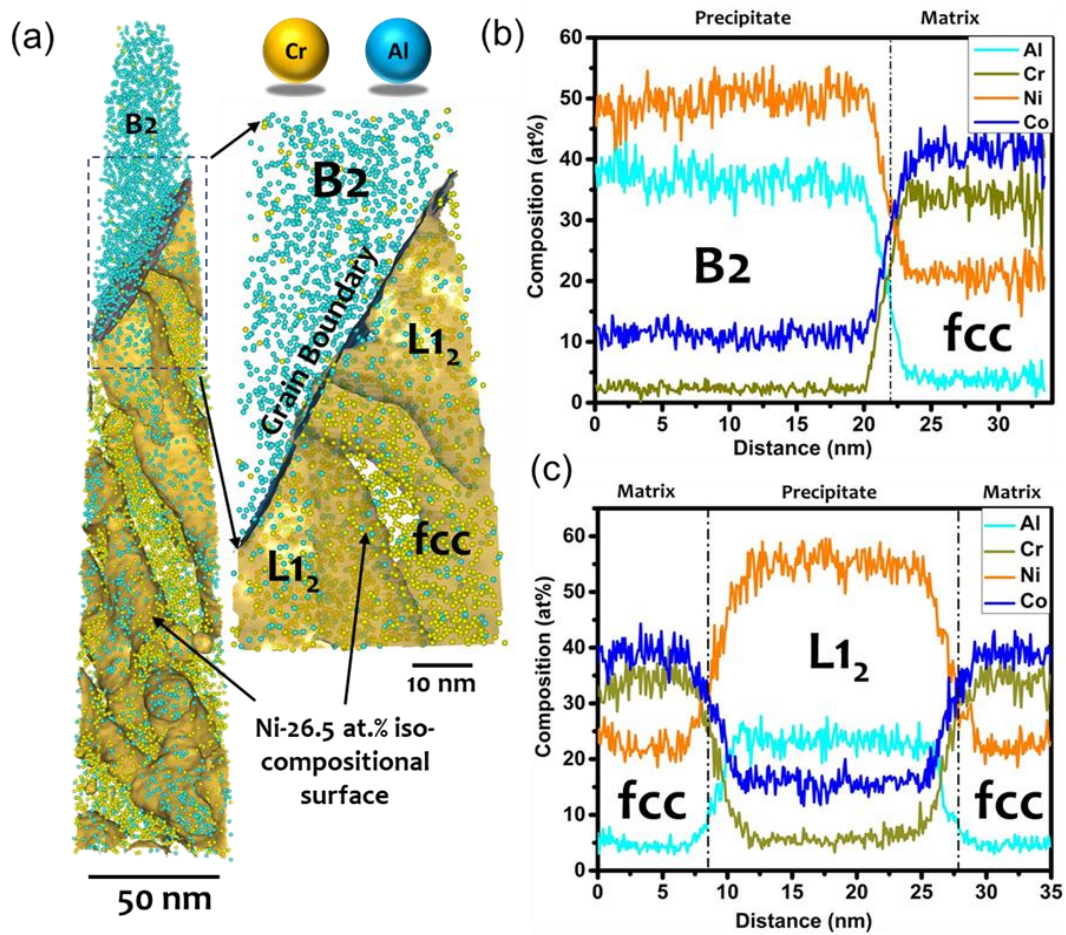

*Figure 3: (a) An APT reconstruction with the distribution of Cr and Al atoms across a B2 precipitate and lamellae structure. Composition profiles across (b) B2 and (f) $L1_2$ precipitates.*

Figure 3(a) shows an atom probe reconstruction of a needle specimen with the distribution of Al (blue) and Cr (golden) atoms taken from a GB region such that both B2 and lamellae structure is captured. The top part of the reconstruction corresponds to B2 precipitate, while the L1$_2$ ordered precipitates are shown by the iso-composition surfaces delineated with a threshold value of 26.5 at.% Ni. The composition profiles across B2 and the adjacent grain fcc matrix (figure 3(b)) reveals a high amount of Ni (~46 at.%) and Al (~40 at.%) with lower content of Co (~11.5 at.%) and Cr (~2.5 at.%) in the precipitate. The diffraction analysis and compositional data indicate these are B2-(Ni,Co,Cr)Al precipitates. The compositional profiles across an L1$_2$ ordered precipitate (figure 3(c)) in the lamellae region show higher content of Ni and Al in the precipitate relative to the adjacent matrix. The compositional values, as shown in Table 1, indicate these are (Ni,Co,Cr)$_3$Al L1$_2$ ordered precipitates. The partitioning coefficient ($K_i$) values across the precipitate-matrix interface for each solute ($i$) are calculated using the relation $K_i = \frac{C_i^P}{C_i^M}$, where $K_i$ is the partitioning coefficient of element $i$. $C_i^P$ and $C_i^M$ are the compositions of element $i$ in the precipitate and matrix. The $K_i$ values are also listed in Table 1. Ni and Al show strong partitioning to the precipitates ($K_{Ni}$=2.4 and $K_{Al}$=4.8), whereas Co and Cr partition to the matrix ($K_{Co}$=0.41, and $K_{Cr}$=0.17). Similar partitioning behavior of these solutes was observed in several Ni-based, Co-based and HEA superalloys responsible for their excellent high-temperature mechanical properties [38–43].

***Table 1:*** *Atomic-scale compositions of matrix and L1$_2$ precipitate after annealing at 50 and 250 hours*

| Element | 50 hours | | | 250 hours | | |
|---|---|---|---|---|---|---|
| | *L1$_2$* | matrix | $K_i$ | *L1$_2$* | matrix | $K_i$ |
| **Ni** | 54.9 | 22.8 | 2.4 | 54.3 | 26 | 2.1 |
| **Co** | 15.92 | 38.5 | 0.41 | 16.4 | 38.5 | 0.42 |
| **Cr** | 5.94 | 33.8 | 0.17 | 7.5 | 29.7 | 0.25 |
| **Al** | 23.23 | 4.8 | 4.8 | 21.7 | 5.8 | 3.74 |

*3.3 High-temperature stability*

To know the stability of these L1$_2$ ordered precipitates, DSC was carried out for NCCA alloy. Figure 4(a) shows the DSC plot during the heating cycle that indicates a slope change near 800 °C that can be attributed to the dissolution of L1$_2$ precipitates in the matrix. Hence, we choose 700 °C for evaluating the stability of the obtained hierarchical microstructure of NCCA alloy by annealing up to 500 hours. Figures 4(b) and (c) show the SEM micrographs after 250 hours and 500 hours of annealing, respectively. Even after 500 hours, the microstructure is stable, i.e., the change in average grain size is not significant (3.4 for 20 hours to 5 µm for 500 hours). In the grain interior, the $L1_2$ precipitates also remain stable without any evidence of decomposition. However, the average lamellae spacing increased from 28 nm for 20 hours to 150 nm for 500 hours annealed sample due to the coarsening of the precipitates. Compositional analysis from a lamellae region (figure 4(d-e)) for the 250 hours sample reveals no significant change in the partitioning behavior of solutes (table 1) across the $L1_2$ precipitate/matrix interface.

Additionally, the average size of B2 precipitates increases from 320 nm (20 hours) to 630 nm (for 250 hours) and remains nearly similar for 500 hours and are also confined only to the GBs, i.e., we couldn't observe them in any of the grain interiors. In contrast, the NCC alloy is fully recrystallized with the same HCA thermomechanical heat treatment (20 hours). This highlights the role B2 precipitates in slowing down the recrystallization kinetics, which results in a partial recrystallized (PR) microstructure for NCCA alloy.

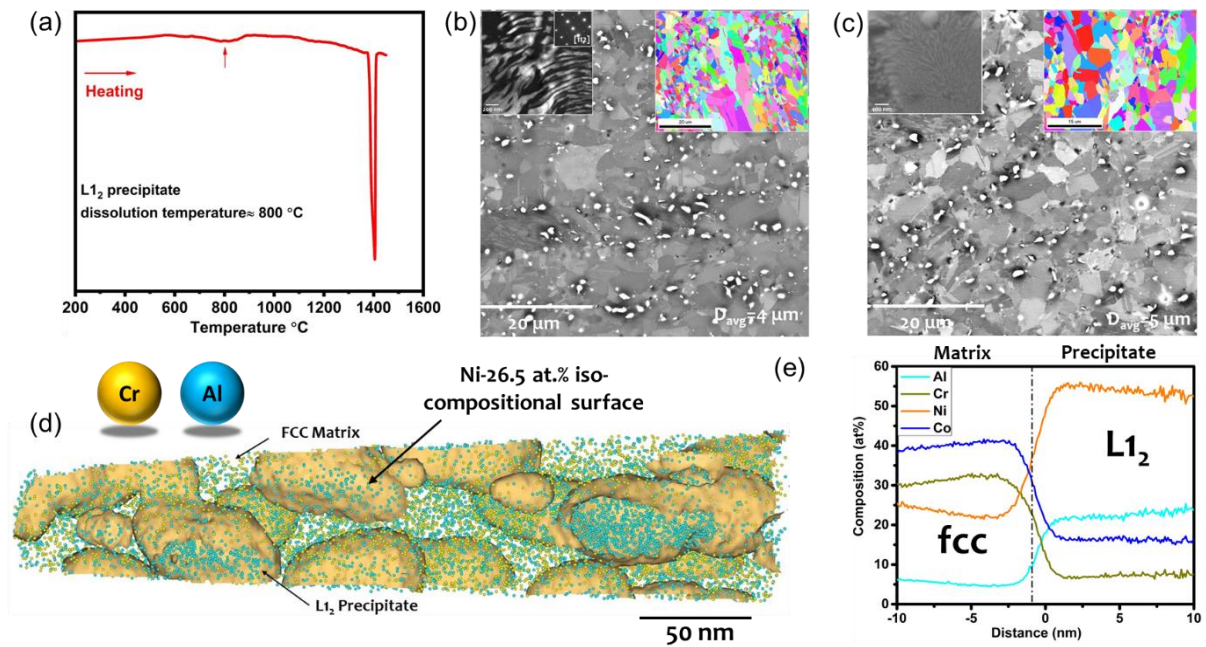

*Figure 4: (a) DSC heating curve for NCCA alloy after HCA thermomechanical heat treatment. BSE microstructures for the NCCA alloy after annealing at 700°C for (b) 250 hours and (c) 500 hours with corresponding IPF maps and lamellae structure inside the grains (see insets). (d) APT reconstruction showing the distribution of Ni and Al atoms across the fcc matrix and $L1_2$ precipitates for NCCA alloy after 250 hours of annealing. (e) Composition profiles across an fcc matrix/$L1_2$ precipitate interface.*

*3.4 Deformation structure*

Figure 5(a) shows the plots between the work hardening rate/true stress vs. true strain for NCC and NCCA alloys. It indicates, NCCA offers a higher work hardening rate within its uniform plastic regime (≤23 % true strain, according to Conside′re instability criteria [44]) illustrated in the same figure. The room temperature tensile deformed NCCA alloy was also investigated by EBSD kernel average misorientation (KAM) map, as shown in figure 5(b). The map shows a high degree of intergranular rotation and development of local misorientations near the GBs, indicating the GBs as major stress concentration centers during deformation. Some regions of the GBs have higher stress concentration (marked by arrows) that is expected to be the B2 precipitate locations. STEM HAADF analysis (figure 5(c-d)) on the deformed sample reveals a significant amount of deformation twinning (see twinning spots in the diffraction pattern taken along [110] zone axis) and the presence of extrinsic stacking faults (ESFs) at the interior of recrystallized grains in the microstructure as shown in a high-resolution HAADF STEM image (figure 5(d)).

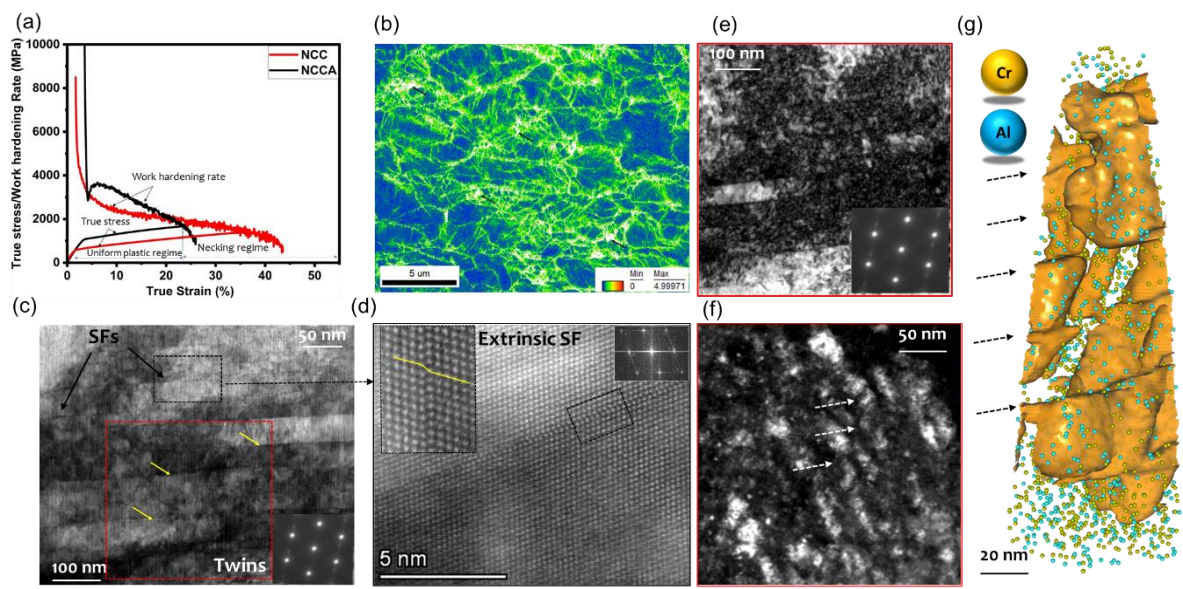

*Figure 5: (a) Comparison of work hardening rate for NCC and NCCA alloys after HCA thermomechanical heat treatment. (b) EBSD KAM map for deformed NCCA alloy after tensile fracture. (c-d) STEM HAADF images from a tensile deformed region showing the presence of deformation twins and extrinsic SFs. (e-f) A pair of BF/DF images (using a superlattice $L1_2$ spot) and an APT reconstruction of a specimen from the same twinned region.*

Figure 5(e-f) shows a pair of TEM brightfield (BF) and DF images from the twinned location marked as red checked box in figure 5(c). The diffraction pattern show twinning spots and also the superlattice spots from $L1_2$ ordering. The DF image taken from one of the superlattice spots

highlight the effect of deformation on the L1$_2$ precipitates. We could see extensive shearing of the precipitates (marked by white arrows) by the movement of dislocations. An APT specimen of the twinned location was prepared from the same TEM lamella. Figure 5(g) shows the APT reconstruction with the distribution of Al and Cr atoms. Ni 22.5 at.% iso-compositional interfaces represented the γ′ precipitates. We observe the cutting of γ′ lamellae and relative displacement of the cut γ′ parts by the shearing events (marked by black arrows). A 3-dimensional visual is also provided as a supplementary video.

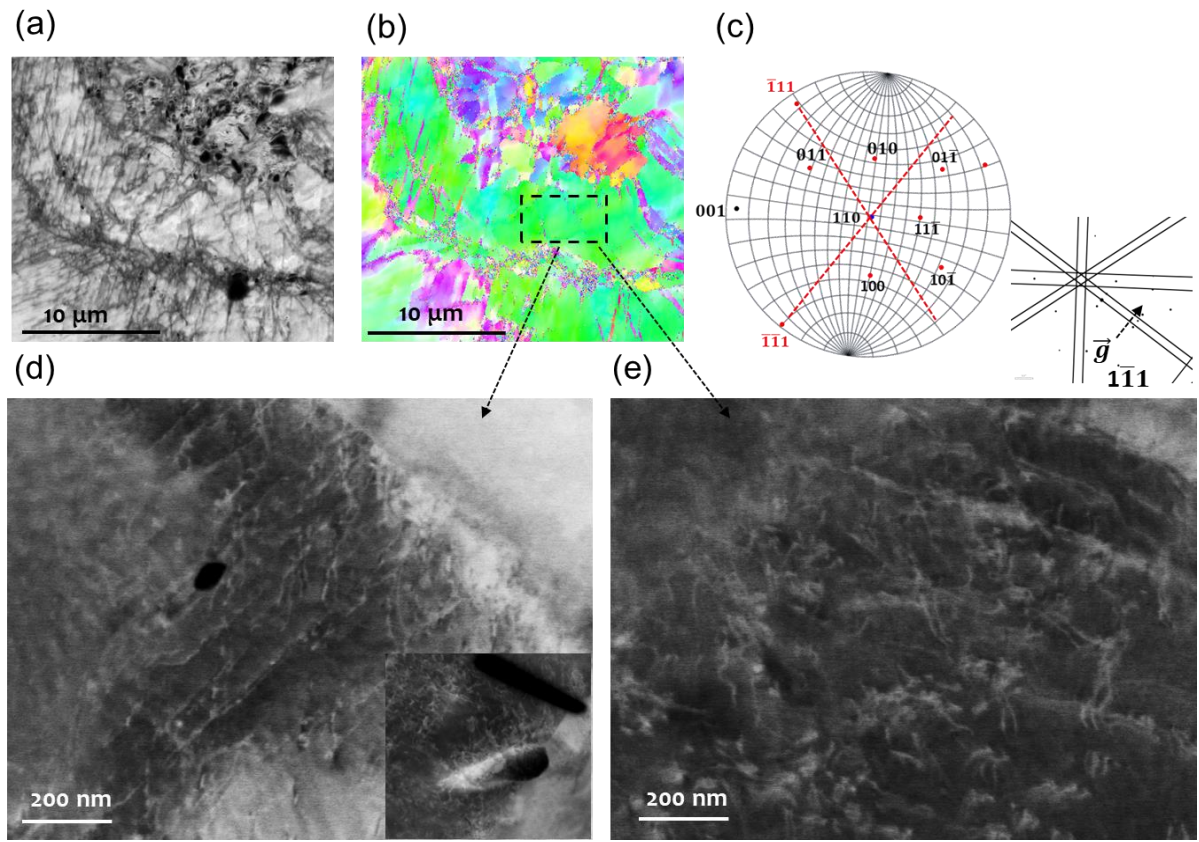

*Figure 6: (a-b) EBSD analysis of a high temperature (670°C) compressive deformed NCCA alloy and (c) corresponding stereogram from an ROI (rectangular checked box) with a simulated ECP pattern in near two-beam conditions (d-e) cECC images taken from the ROI showing strong channeling contrast with the defect features appearing brighter.*

cECCI was carried out for the sample that was compressively deformed at 670°C. A deformed region was located with an orientation close to the [110] zone axis. A near two-beam condition for the ROI was achieved with $g = 1\bar{1}1$ (where g is the diffraction vector) by tilting and rotation of the sample stage in SEM (see the experimental section for details). The corresponding stereogram is also shown in figure 2(c). Figure 6(d-e) shows the recorded cECC images from the same ROI. The images clearly show channeling contrast from the defects that appear brighter to the dark surrounding regions. We observe sub-structure boundaries whose traces lie on {111} planes. Additionally, the dislocations are piled at these sub-structure boundaries and around the B2 precipitate (see inset in figure 6(d)).

**4. Discussion**

The above experiments highlight the sensitivity of an MEA towards alloying (Al addition) and thermomechanical processing (HCA) that led to enhanced microstructural stability and mechanical properties at room and high temperatures. More specifically, we demonstrated the formation of a hierarchical microstructure comprising soft recrystallized zones with $L1_2$ ordered strengthening precipitates, hard non-recrystallized zones, and B2 ordered precipitates at GBs whose compound effect resulted in a strong, tough, and high-temperature stable MEA. Now we will discuss their contributions.

*4.1 Recrystallization kinetics – Role of B2 precipitates*

The cold-worked state is thermodynamically unstable with high stored energy in the form of lattice strains created by the generation and accumulation of many dislocations and point defects. The high stored energy is released via three successive processes that are recovery, recrystallization, and grain coarsening during annealing at high temperatures [45,46]. Recovery involves point defect elimination and rearrangement of dislocations known as polygonization resulting in low-angle sub-grain boundaries. Some of these recovered sub-grain regions that are highly misoriented to the adjacent deformed regions evolve into new grains with high angle boundaries (annealing induced), i.e., recrystallization. More specifically, the sub-grains act as sites for the nucleation of recrystallized grains. This nucleation mechanism is different from the kinetic model of nucleation applied in solidification or solid-state phase transformations where the atoms have to build up to a critical size with new structures different from the parent structure [47,48]. While in recrystallization, the new grains follow the same structure and orientation as present in the deformed state [45]. Hence, it is expected that the formation of new grains occurs at a much faster rate.

Given this background, now we discuss the sequence of microstructural development during annealing in the present NCCA alloy. Figure 7(a) shows a BSE image of a non-recrystallized grain with large amount of deformation-induced planar slips. We could also observe a few B2 precipitates inside the grain without any evidence of recrystallization around them. This indicates the system prefers earlier nucleation of B2 rather than new grains, as also observed in earlier studies [46]. These precipitates are incoherent and might be heterogeneously promoted to form on the highly deformed and misoriented sites in the non-recrystallized grains [49,50]. On further annealing, these B2 precipitates grow and induce additional volumetric strain around the precipitates in the surrounding grain. This creates additional strain and orientation gradients that can act as nucleation sites for the new recrystallized grains, figure 7(b). The is also evidenced from the KAM map (figure 7(c) and EBSD IPF map (figure 7(d)) of a similar region from a non-recrystallized grain containing few B2 precipitates. The KAM map shows a relatively higher strain around the B2 precipitates to the surrounding grain. The EBSD IPF map reveals nucleation and growth of a recrystallized new grain with negligible strain inside. The B2 precipitates also pins the GB. This nucleation event is referred to as particle stimulated nucleation (PSN) of new grain [46]. Hence, the distribution of B2 precipitates in the non-recrystallization grains is proposed to control and promote finer size distribution of the new recrystallized grains.

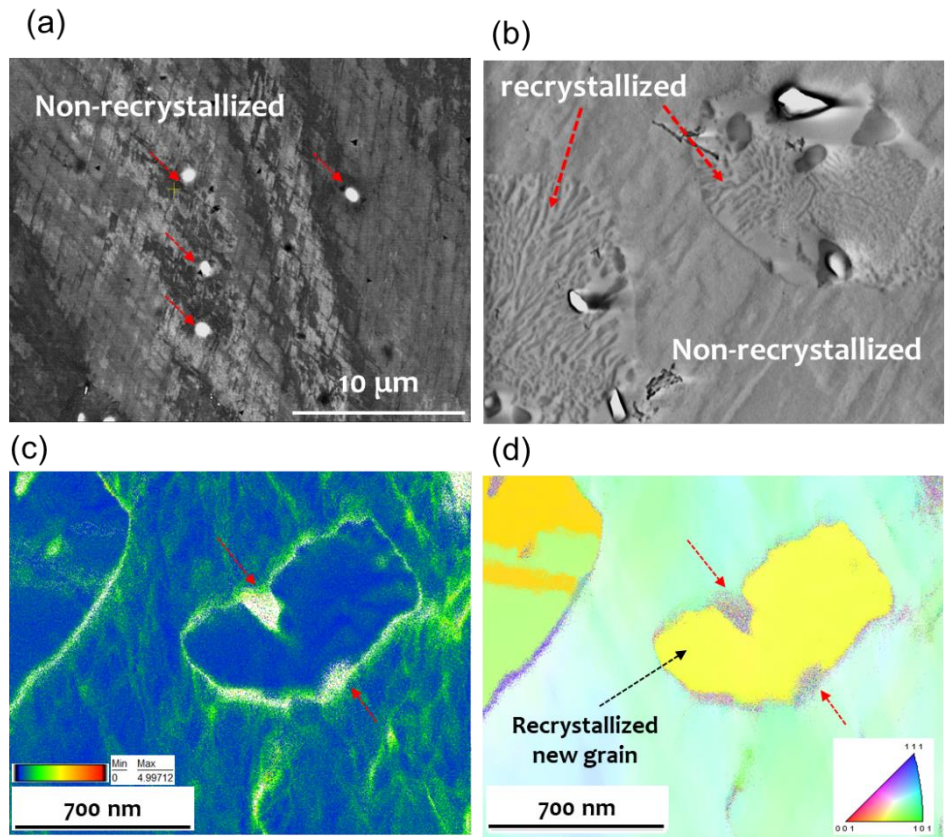

*Figure 7:* (a) A BSE image of NCCA alloy after HCA thermomechanical heat treatment from a non-recrystallized grain containing bright B2 precipitates. (b) Formation of new recrystallized grains in the vicinity of B2 precipitates (c) EBSD KAM map centered on a few B2 precipitates and (d) EBSD IPF map from the same region showing nucleation/growth and pinning of a new recrystallized grain by B2 precipitates.

Further, when the grain boundaries (GBs) of the nucleated new grains migrate, simultaneously, discontinuous precipitation [37] of strengthening coherent L1$_2$ ordered regions takes place in the form of lamella structure, as we see from the microstructures shown in figure 2(b and e). The fineness of the lamella structure directly depends on the size of the recrystallized new grains, and hence we observe a very dense distribution of lamellae structure across the recrystallized regions of the microstructure. Similar discontinuous precipitation of L1$_2$ ordered regions was observed in FeNiCoCrAl HEA during recrystallization [24,51]. Interestingly, we made a new observation that further deepened the understanding of the role of B2 on the microstructure, Figure 8(a) shows the top section of the APT reconstruction (figure 3(a)) centered on the GB. We plotted composition profiles across B2/GB/L1$_2$ and B2/GB/fcc matrix, as shown in figure 8(b). The former profile reveals a significant amount of segregation of Ni (up to 63 at.%) at the GB. In comparison, Co and Cr are depleted by up to 8 at.% and 1.5 at.%, respectively, as compared to the adjacent compositions of B2 and L1$_2$ precipitate. However, we couldn't notice any similar segregation of Ni or depletion of Co/Cr at the GB between B2 and adjacent grain fcc matrix. Figure 8(c) shows the same reconstruction on xz plane viewed from the bottom of the GB that clearly distinguishes L1$_2$ and fcc matrix. We plotted a 2D Ni compositional map of the GB plane for the Ni distribution. We observed in-homogenous distribution, i.e., the segregation of Ni is confined only to the locations where L1$_2$ precipitates are nucleated. This might directly impact the growth and coarsening of B2 precipitates with the annealing time and hence the alloy microstructure. More specifically, from these atomic-scale observations, we propose a rate-limiting diffusional mechanism that is demonstrated through a schematic in figure 8(d).

As annealing progresses, the alloy microstructure coarsens, i.e., growth/coarsening of B2 precipitates, recrystallized grains, and hence the lamella structure. In the left schematic diagram, we have also kept the average concentration (calculated from the composition profiles) of all the solutes in the respective phases. During annealing, the B2 precipitate interface advances to the lamella region and hence needs local diffusion or rearrangement of atoms to obtain the equilibrium B2 composition. We have marked as **zone 1** for L1$_2$ and **zone 2** for fcc matrix regions in the schematic. The solute diffusion directions are also indicated from the relative solute concentrations and observed GB segregation behavior. When the interface of B2 advances, say by $x$ distance, Ni will be higher in zone 1 while lower in zone 2 to the required for equilibrium concentration in B2. Thus, the excess Ni from zone 1 needs to diffuse to zone 2. The evidence of segregation of Ni at the GB between B2 and L1$_2$ indicates that the excess Ni will diffuse along the GB to zone 2. The diffusion will continue to happen until it reaches the required equilibrium Ni concentration in B2. Parallelly, since Al is lower in zone 1 and zone 2 as compared to B2, the Al will be consumed from the lamellar structure to reach the equilibrium Al concentration in B2.

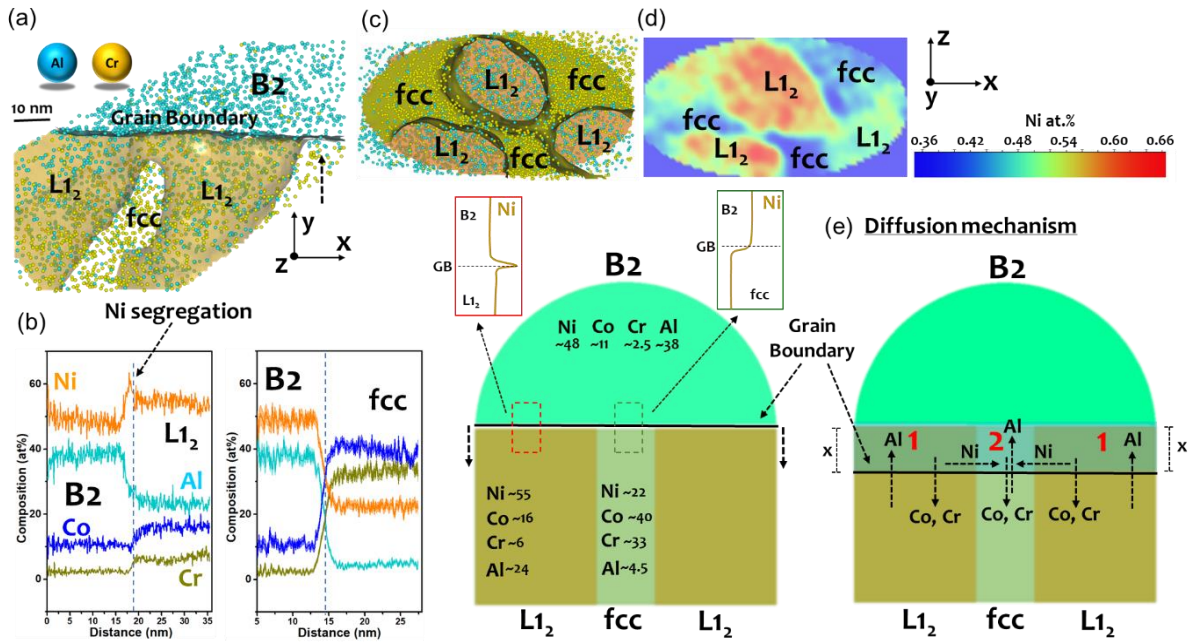

*Figure 8: (a) Top section of the APT reconstruction shown in figure 3(a) centered on the GB. (b) Compositional profiles across B2/GB/L1$_2$ and B2/GB/fcc matrix. (c) A 2D section of the reconstruction on xz plane viewed from the bottom of the GB (see dashed arrow). (d) A 2D compositional map of Ni on the GB plane viewed on the same xz plane and respective color-composition scale. (e) A schematic illustration of solute diffusion during the growth/coarsening of B2 precipitate.*

Further, both Co and Cr are higher in zone 1 and 2 than the required in B2; hence, they will diffuse away from GB towards the lamellae structure. This is supported by our observation of Co, and Cr depletion near the GB shared between B2 and L1$_2$. Overall, this diffusional scheme will result in the growth of the B2 precipitate at the expense of the lamellar structure. The segregation of Ni at GB between B2 and L1$_2$ indicates its diffusion as a rate-limiting for the growth of the B2 precipitates and hence acts as a critical factor for the microstructural stability of the alloy. We observe slower recrystallization kinetics and enhanced microstructural and mechanical stability of NCCA alloy at 700°C up to 500 hours, indicating Ni might be the slow diffusing compared to other elements and rate-controlling solute. This is supported by recent work on NiCoCrFe MEA, where it was shown by detailed diffusional experiments that Ni is the slowest diffusing species [52]. Quantitatively, the slower recrystallization kinetics can be explained by considering the migration velocity $v$ of a GB during recrystallization [53,54], and it is given as $v = m \times P$, where *m* is interfacial mobility, and *P* is net driving pressure acting on the GB. Considering Zener drag ($P_Z$) [55,56] and solute drag ($P_S$) [57,58] effects, the net driving pressure becomes: $P = P_D - P_Z - P_S$. Where, $P_D$ is the driving pressure provided by the elastic energy stored through the dislocation content $\rho$ [59] that is given by $P_D = \frac{\rho G b^2}{2}$. The $\rho$ value is considered to be $5 \times 10^{14} \, m^{-2}$, for a typical cold-worked material. Shear modulus (G) is assumed to be 87 GPa for a NiCoCr alloy [60]. $b = 0.252$ nm is the length of the Burgers vector. Hence, $P_D$ is estimated to be $1.3 \times 10^6 \, J/m^2$. Zener drag is caused by the B2 ordered precipitates at the GBs and can be estimated by $P_z = 3f\gamma_B/2r$, where *f* = 0.08, and *r* = 160 nm is the corresponding area fraction, and the average equivalent radius of B2 ordered precipitates. $\gamma_B$ is the GB energy, assumed to be 0.625 J/m² [61]. The calculated Zener drag pressure is $4.6 \times 10^5 J/m^2$, that results in the reduction of the driving pressure for static recrystallization by ~34 %. The solute drag pressure is calculated by $P_S = \frac{\alpha c v}{(1+\beta^2 v^2)}$. Where $\alpha$ and $\beta$ are the parameters related to interaction energy $E_x$ and effective diffusion of solutes $D_{eff}$ along GBs, and *c* is bulk solute concentration. The studied alloy is a multi-component system, and it isn't easy to estimate the value of $E_x$ and $D_{eff}$. However, the solute drag force $P_s$ is inversely proportional to the $D_{eff}$. Since we found that the diffusivity of Ni is rate controlling and slowest among other elements in the alloy, it increases the solute drag pressure $P_S$ that contributes to reducing the net driving pressure on the GBs and lowers the recrystallization kinetics i.e., the GBs migrate at a slower rate. This effect will directly influence the growth/coarsening of L1$_2$ precipitates since the lamellae structure is formed via discontinuous precipitation due to GB migration [37]. The top part of the APT reconstruction (figure 3(a)) clearly shows a GB pinned by a B2 precipitate and discontinuous precipitation of L1$_2$ precipitates from the same GB below.

*4.2 Phase stability of strengthening L1$_2$ precipitates*

Another important observation is that there is no evidence of L1$_2$ phase decomposition inside the recrystallization grains even after 500 hours of annealing at 700°C. However, in Fe containing HEAs, the L1$_2$ phase is unstable beyond 550°C, and it decomposes to body-centered-cubic (bcc) ordered B2 phase inside the grains. Fe is well known for destabilizing L1$_2$ ordered structures in Ni-base alloys. Ma et al. [62] showed by experiments that by increasing Fe content (1 to 10 wt.%) in Ni-33Cr-10Al based alloy, the L1$_2$ structure phase starts disappearing, and other phases (bcc (β) and fcc (γ)) evolve in the microstructure. Thermodynamic Calculations (ThermoCal) supported the observations and revealed shrinking of γ′ phase field and replacement by γ + β phase-field as the Fe content increases in the alloy. This was attributed to the lowering of γ and β Gibbs free energies with the increase in Fe while Gibbs free energy of γ′ escalates that lowers the total Gibbs free energy by the formation of γ + β phases. Similar destabilizing of L1$_2$ structure by Fe addition is reported in a non-equiatomic NiCoCrFeAl/Ti HEA/concentrated alloys promoting other undesirable phases [51,63]. It was also reported that doping of Fe into Ni$_3$Al L1$_2$ ordered lattice reduces the order-disorder transition temperature by ~ 180°C that directly relates to the lowering of their phase stability [64]. Hence, removal of Fe in the NiCoCrFeAlTi concentrated alloy led to an increase in the dissolution temperature by 90°C and improved the high temperature coarsening resistance of L1$_2$ precipitates [63]. In the present MEA, since Fe is not present, it is expected the decomposition of L1$_2$ precipitates is prevented, as we also observed experimentally.

*4.3 Deformation structure and individual contributions to strengthening*

We observe a higher strain/work hardening rate for NCCA. The tensile deformed/fractured microstructure reveals extensive twinning and the presence of SFs across the lamella structure. Even in NCC MEA, deformation twinning and SFs are seen in tensile deformed/fractured conditions and hence show a high degree of strain hardening. The tendency of twinning and formation of SFs is directly related to their low stacking fault energy (SFE ~ 18 to 22 mJ/m$^2$) [12,65] that promotes easy dissociation of perfect dislocations into a/6<112> partial dislocations. The twinning and SFs acts as strong barriers to the movement of dislocations and hence increases the strain hardening rates greater than several HEAs with higher SFEs. In the deformed/fractured NCCA alloy, we observe additional shearing of L1$_2$ ordered regions by the movement of dislocations and excessive local misorientations (figure 5) near the B2 precipitates. Hence, these combined factors induce a higher work hardening rate for NCCA alloy. cECCI images from the high temperature compressive deformed NCCA alloy reveal excessive sub-structure formation and dislocation entanglement around them, GBs and B2 precipitates indicating their strong contribution in resisting deformation even at high temperatures.

To estimate their contributions in NCCA alloy, a strengthening model is proposed by considering the heterogeneity in the microstructure. The rule of the mixture was used, i.e., $\sigma_y = f_{RZ}\Delta\sigma_{RZ} + (1 - f_{RZ})\Delta\sigma_{NRZ}$. Where $f_{RZ}$ is a fraction of soft recrystallized zones, $\Delta\sigma_{RZ}$ and $\Delta\sigma_{NRZ}$ are the 0.2% YS for the recrystallized and non-recrystallized zones, respectively. $\Delta\sigma_{NRZ}$ value can be approximated by considering the yield strength of cold-rolled NCCA alloy (graph shown in supplementary Fig. S1), around 1.1 GPa. $\Delta\sigma_{RZ}$ is estimating by the sum of the contribution from solid solution strengthening, grain size strengthening, and precipitation strengthening, expressed as $\Delta\sigma_{RZ} = \Delta\sigma_s + \Delta\sigma_g + \Delta\sigma_p$. The contribution from solid solution strengthening is determined experimentally by conducting tensile tests for homogenized NCCA and NCC alloys. The corresponding tensile 0.2% YS values for NCCA and NCC alloys are 335 MPa and 250 MPa, respectively (see supplementary Fig. S(2) for tensile stress-strain curves). Thus $\Delta\sigma_s = \sigma_{NCCA} - \sigma_{NCC}$ and it is estimated to be 85 MPa. The average grain size of NCCA alloy after annealing is 3.4 μm. Based on the Hall-Petch relation, $\Delta\sigma_g = \sigma_0 + \frac{k}{\sqrt{d}}$ where, $\sigma_0$ is frictional stress with a value of 154 MPa, and $k$ is the Hall-Petch coefficient, i.e., 684 MPa$\sqrt{\mu m}$, adapted from the literature (derived for similar kinds of alloy) [66]. Thus $\Delta\sigma_g$ is calculated to be 525 MPa. Contribution from the precipitates has two components, i.e., due to L1$_2$ and B2 ordered precipitates. Here we assume precipitate shearing by dislocations for the former and Orowan looping for the latter. Strengthening by precipitate shearing can be estimated using the equation $\Delta\sigma_{ps} = M 0.81 \frac{\gamma_{APB}}{2b} \sqrt{\left(\frac{3\pi f}{8}\right)}$ where $M$ is Taylor factor (3.06 for a polycrystalline matrix, used to convert shear stress value to flow stress) [67], $b$ is the magnitude of Burgers vector of the matrix (0.252), $f$ is L1$_2$ precipitate area fraction obtained to be ~ 30%. $\gamma_{APB}$ is the anti-phase boundary energy (APB) of L1$_2$ precipitates considered to be 130 mJ/m$^2$ [68,69]. The estimated contribution from the L1$_2$ precipitates is $\sigma_{ps} = 379$ MPa. Similarly, the contribution from B2 precipitates can be estimated by the equation $\Delta\sigma_{Po} = M \cdot \frac{0.4 \times Gb}{\pi\sqrt{1-v}} \cdot \frac{\ln(2\tilde{r}/b)}{L_p}$, where $G$ is shear modulus ~ 87 GPa (assuming similar to CoCrNi), $v = 0.3$ is the Poisson's ratio, $\tilde{r} = \sqrt{\frac{2}{3}} r$, where r is the mean particle radius in the slip plane, and $L_p \approx 1000\ nm$ is mean inter-particle distance. Thus, the contribution from the B2 order precipitates is 71 MPa. The combined contribution from $\Delta\sigma_s$, $\Delta\sigma_g$, $\Delta\sigma_p$, is $\Delta\sigma_{RZ} \sim 1060$ MPa. Thus, 0.2% YS of NCCA alloy (including $\Delta\sigma_{NRZ}$) can be estimated as $\sigma_y = 1072$ MPa. The value is near to the experimentally determined 0.2% YS value of 1030 MPa.

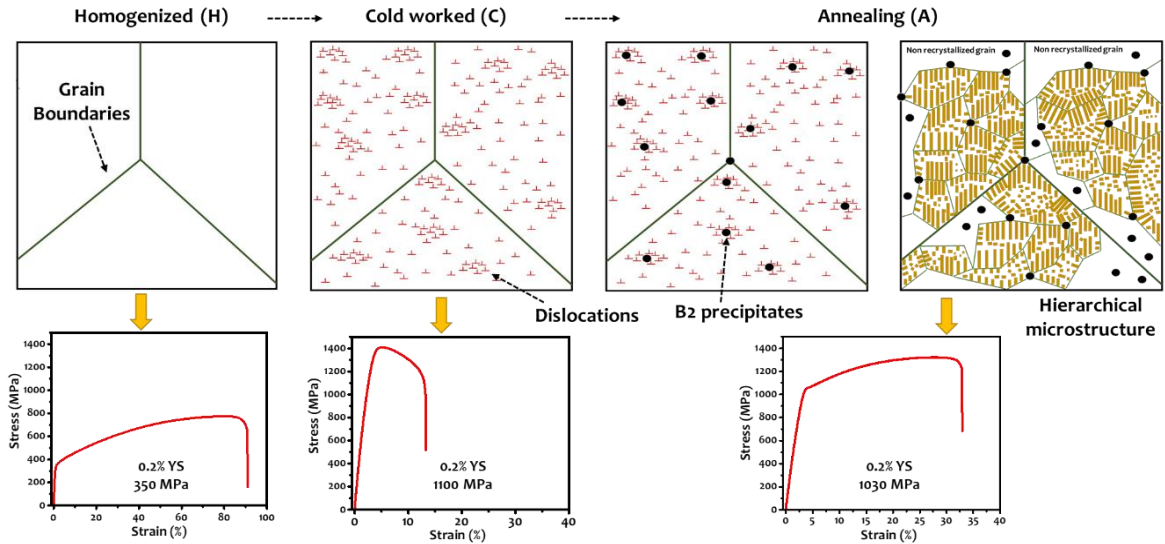

*Figure 9: A schematic illustration of the sequence of microstructural evolution of NCCA alloy during HCA thermomechanical heat treatment with the corresponding tensile stress vs strain plots.*

## 5. Conclusions

In conclusion, NiCoCr MEA alloy offers enormous scope for developing structural alloys with promising mechanical properties under a wide temperature range, i.e., from cryogenic to high temperatures. The high sensitivity of the alloy towards alloying (Al) and thermo-mechanical processing (PR) resulted in hierarchical microstructure consists of soft recrystallized zones, coarse non-recrystallized zones, and ordered precipitates ($L1_2$ and B2). The sequence of microstructural development is illustrated schematically in figure 9. The composite microstructure displays a tensile 0.2% YS of 1030 MPa, UTS of 1320 MPa with 32 % ductility at room temperature, and retains up to 910 MPa at 670°C. The microstructure is also stable at 700°C up to 500 hours. We observe a critical role of B2 precipitates in lowering down the recrystallization kinetics. Atomic-scale compositional analysis reveals Ni as the rate-limiting solute in the alloy that controls the growth/coarsening of B2 precipitates and can directly impact the size distribution of recrystallized grains and hence the stability of strengthening lamellae structure.

Deformed microstructures reveal extensive twinning and formation of stacking faults, shearing of $L1_2$ precipitates by dislocations, and strain accumulation at the GBs and B2 precipitates that resulted in higher strain hardening for NCCA alloy as compared to NCC alloy.

Future directions will be to investigate in detail the deformation structure of non-recrystallized in the present MEA that promotes the nucleation of B2 and the deformation behavior of NCCA alloy at room and high temperatures. Since the dissolution temperature of strengthening $L1_2$ precipitates is 800°C, a selection of further alloying such as Ta and Ti knew to stabilize the L1$_2$ ordered structure can improve the PR MEA microstructural stability at higher temperatures. Thus, it opens new possibilities for designing high-performance MEAs.

**Acknowledgments**

The authors gratefully acknowledge the Advanced Facility for Microscopy and Microanalysis (AFMM), Indian Institute of Science, Bangalore, for providing access to TEM and APT facilities. NB is grateful to SERB India for National Postdoc Fellowship (NPDF). SKM acknowledges the financial support from IISc for SEED Grant and MPG-IISc Partner Group.